\documentclass[twocolumn,showpacs,preprintnumbers,superscriptaddress,amsmath,amssymb,prl]{revtex4}

\usepackage{graphicx}
\usepackage{dcolumn}
\usepackage{bm}
\usepackage{amsmath}
\newcommand{\be}{\begin{equation}}
\newcommand{\ee}{\end{equation}}
\newcommand{\ba}{\begin{eqnarray}}
\newcommand{\ea}{\end{eqnarray}}

\begin{document}

\preprint{APS preprint}

\title{Power Law Distributions of Offspring and Generation Numbers in
Branching Models of Earthquake Triggering}

\author{A. Saichev}
\affiliation{Mathematical Department,
Nizhny Novgorod State University, Gagarin prosp. 23,
Nizhny Novgorod, 603950, Russia}
\affiliation{Institute of Geophysics and Planetary Physics,
University of California, Los Angeles, CA 90095}

\author{A. Helmstetter}
\affiliation{Institute of Geophysics and Planetary Physics,
University of California, Los Angeles, CA 90095}

\author{D. Sornette}
\affiliation{Institute of Geophysics and Planetary Physics,
University of California, Los Angeles, CA 90095}
\affiliation{Department of Earth and Space Sciences, University of
California, Los Angeles, CA 90095\label{ess}}
\affiliation{Laboratoire de Physique de la Mati\`ere Condens\'ee,
CNRS UMR 6622 and Universit\'e de Nice-Sophia Antipolis, 06108
Nice Cedex 2, France}
\email{sornette@moho.ess.ucla.edu}

\date{\today}

\begin{abstract}
We consider a general stochastic branching process,
which is relevant to earthquakes as well as to many other systems, and 
we study the distributions of the total number of offsprings (direct and
indirect aftershocks in seismicity) and of the total number of
generations before extinction. We apply our results to 
a branching model of triggered seismicity, the
ETAS (epidemic-type aftershock sequence) model.
The ETAS model assumes that each earthquake can trigger
other earthquakes (``aftershocks''). An aftershock sequence
results in this model from the cascade of aftershocks of each past earthquake.
Due to the large fluctuations of the number of aftershocks triggered
directly by any earthquake (``fertility''), there is a large
variability of the total number of aftershocks from one sequence to
another, for the same mainshock magnitude. 
We study the regime where the 
distribution of fertilities $\mu$ is characterized 
by a power law $\sim 1/\mu^{1+\gamma}$. For earthquakes, we expect
such a power-distribution of fertilities with $\gamma = b/\alpha$ based 
on the Gutenberg-Richter magnitude distribution $ \sim 10^{-bm}$ and on
the increase $\sim 10^{\alpha m}$ of the number of aftershocks with
the mainshock magnitude $m$.  We derive the asymptotic distributions 
$p_r(r)$ and $p_g(g)$ of the total number $r$ of offsprings and of the 
total number $g$ of generations until extinction following a mainshock. 
In the regime $\gamma<2$ for which the distribution of fertilities has
an infinite variance, we find  $p_r(r) \sim 1/r^{1+{1 \over \gamma}}$ and
$p_g(g) \sim 1/g^{1+{1 \over \gamma -1}}$. This should be compared
with the distributions $p_r(r) \sim 1/r^{1+{1 \over 2}}$ and
$p_g(g) \sim 1/g^{1+{1}}$ obtained for standard branching processes with
finite variance. These predictions are checked by numerical
simulations. Our results apply directly to the ETAS model whose prefered
values $\alpha=0.8$ and $b=1$ puts it in the regime
where the distribution of fertilities has an infinite variance. 
More generally, our results apply to any stochastic branching process
with a power-law distribution of offsprings per mother.
\end{abstract}


\maketitle

\section{Introduction}

All large earthquakes are followed by an increase of seismic activity 
known as ``aftershocks.'' Aftershock sequences of small earthquakes are less 
obvious because the aftershock productivity is weaker, but can be clearly
observed when stacking many sequences \cite{alpha}. It is thus natural 
to assume that each earthquake can trigger its own aftershock sequence,
and that observed aftershock sequences result from the cascade of direct
aftershocks (triggered directly by the mainshock) and indirect aftershocks
(triggered by a previous aftershock of the mainshock). This assumption is 
the basis of the Epidemic Type Aftershock Sequence model (ETAS) 
of seismicity \cite{Ogata88,Ogata99,KK81,KK87,helmsor1}, which 
describes earthquake triggering as a branching process. In addition, 
the ETAS model includes the Omori law decay $ \sim 1/(t+c)^p$
of the number of direct aftershocks as a function of the time $t$ since the mainshock,
where $c$ is a small constant and $p$ is an exponent close to but larger than 1.
Previous works on this model have shown that the ETAS model provides a better fit
to aftershock sequences than a single Omori law (no secondary aftershocks)
 \cite{GuoO97} and that a significant fraction of aftershocks, of the order of 
80\%, are secondary aftershocks \cite{Felzer1,Felzer2}.
The ETAS model has been used in many studies \cite{KK81,Felzer1} to describe or predict
the spatio-temporal distribution of seismicity and reproduces
many properties of real seismicity, including a renormalization
of the Omori exponent from the local Omori law  (direct aftershocks) to the
global Omori law  (observed rate of aftershocks including secondary aftershocks) 
\cite{Ogata99,SorSor,helmsor1}, B{\aa}th's law \cite{Bathlawpap}, a diffusion
 of aftershocks and realistic foreshock properties \cite{foreshockp,Forexp}.
In this work, we present an analytical derivation of the distribution
of the total number of aftershocks, summed over all generations
of the cascade of aftershock triggering, and of the distribution of
the total number of generations of aftershocks before
extinction.

There are two well known statistical laws that describe the scale-invariance of 
earthquake physics with respect to magnitudes.
First,  the (complementary cumulative) Gutenberg-Richter (GR)
distribution of earthquake magnitudes gives the probability 
\be
p_m(m) \sim 10^{-b m}
\label{GR}
\ee
that an earthquake has
magnitude equal to or larger than $m$.
This magnitude distribution $p_m(m)$ is not dependent
on the magnitude of the triggering earthquake, i.e., a large
earthquake can be triggered by a smaller one \cite{alpha,Forexp}.

Second, the average number of aftershocks triggered directly by an earthquake 
of magnitude $m$ is found to increase with $m$ as
\be
N_m = K ~10^{\alpha m}
\label{alpha}
\ee
where $K$ is a numerical factor independent of the magnitude.
The number of direct aftershocks cannot be measured, because what is observed
is the total number of direct and secondary aftershocks.
If earthquake triggering can be described by a branching process 
such as the ETAS model, then it can be shown that the scaling of the total 
number of aftershocks with the mainshock magnitude also obeys the law
(\ref{alpha}), but with a larger factor $K'$ which accounts for the
cascades of secondary aftershocks \cite{helmsor1}. 
The exponent $\alpha$ in (\ref{alpha})
can thus be measured from a fit of the total number of aftershocks with the 
mainshock magnitude. Fits of the total number of aftershocks
as a function of the mainshock magnitude in individual sequences
support (\ref{alpha}) with an exponent $\alpha$ in the range 0.75-1
\cite{6,9,11,13}. However, the precision of these studies is 
limited by the narrow range of mainshock magnitudes considered and the
large scatter of the number of aftershocks per mainshock. 
The value of $\alpha$ estimated in these studies may also be biased by the 
arbitrary constraint that aftershocks must be smaller than the mainshock,
by the incompleteness of the catalog just after the mainshock, and by the 
background seismicity at large times after the mainshock. 
Other studies have measured the exponent $\alpha$ in (\ref{alpha})
by calibrating the ETAS model to real data (individual aftershock sequences 
or complete catalogs) using maximum likelihood methods \cite{Ogata88,K91,GuoO97,Zhuangetal}. 
These studies found a large scatter of the $\alpha$ exponent in the range 
$0.2-2$. It is not yet clear if this range of values reflects a real 
variability of $\alpha$ or the inaccuracy of the estimation of $\alpha$.
One of us used a stacking method to estimate the
average rate of earthquakes triggered (directly or indirectly)
by a previous earthquake as a function of the magnitude of the triggering 
earthquake \cite{alpha}, without constrain on the aftershock magnitude.
For the catalog of the Southern California Data Center for Southern California, 
using the time period 1975-2003 and $m\geq 3 $ earthquakes, $\alpha$ is found 
equal to $0.8\pm0.1$, smaller than $b=1 \pm 0.1$.
Small earthquakes are thus collectively more important than larger earthquakes
for earthquake triggering  if $\alpha \leq b$, because they are much more 
numerous than larger ones.

Let us combine the two laws (\ref{GR}) and (\ref{alpha}) to get the
unconditional probability
density for the number $N_m$ of events triggered  directly by any event
whose magnitude $m$ is drawn at random from the GR law. For this, we note that
\be
{\rm Pr}({\rm fertility} \geq N_m) = 
{\rm Pr}(K ~10^{\alpha m} \geq N_m) = 
{\rm Pr}(m \geq {1 \over \alpha} \log_{10}{N_m \over K}) 
\sim 10^{-{b \over \alpha} \log_{10}{N_m \over K}} \sim {1 \over N_m^{b / \alpha}},
\label{pojid}
\ee
where $Pr$ means ``Probability''.
The first equality makes use of (\ref{alpha}) and the third
equivalence makes use of (\ref{GR}). Hence, the probability density of
the fertilities $N_m$, for a magnitude $m$ drawn at random in the GR law, is
\be
p_{N_m}(N_m) \sim {1  \over {N_m}^{1+\gamma}}~, 
\label{aerapri}
\ee
with $\gamma = b/\alpha$. Typically, $b=1$ and $\alpha \approx 0.8$ \cite{alpha}
leading to $\gamma \approx 1.25$. Because $\gamma < 2$, the variance
of the fertility, 
for an earthquake of arbitrary magnitude drawn from the GR law, 
is mathematically infinite (if we assume the 
absence of a roll-off in the GR distribution, see below).

Beyond earthquakes, a multitude of phenomena can also be described by
branching processes with power-law distributions of fertility.
Stochastic branching processes indeed describe well a multitude of
phenomena \cite{Athreya,Sankaranarayanan} from chain reactions in nuclear 
and particle physics, material rupture, fragmentation and earthquake 
processes, critical percolation cluster sizes and population growth models,
to population and biological dynamics, epidemics,
economic and social cascades and so on. Branching processes
are also of particular interest because deep connections have
been established with critical phenomena \cite{Vere76,Vere77}.
Epidemic transmission of diseases, and more generally transmission 
processes involving avalanches spreading on networks such as the
World Wide Web, cellular metabolic network, ecological food webs, 
social networks, and so on exhibit such heavy-tail 
probability density functions (PDF)
given by (\ref{aera}) below, as a consequence of the well-documented power law 
distribution of connectivities among nodes. Our results are thus relevant
to systems in which the number of offsprings may be large due to
long-range interactions, long-memory effects or large deviation processes.
Goh et al. \cite{Goh} actually derive results that overlap with ours
in the context of avalanches in social networks.

In branching processes with a finite variance of the number of daughters 
per mothers, various quantities exhibit power law distributions 
with universal exponents at criticality (in statistical physics,
the term ``universal'' refers to the independence of the critical 
exponents on the microscopic details of the physics). This includes
the distributions of cluster sizes, of the number of
generations before extinction and of durations which are mean field 
\cite{Athreya,Sankaranarayanan} (``mean field'' 
refers to the branching approximation which leads to a lack of 
dependence on the space dimension).
In the case of earthquakes and for other systems mentioned above,
the distribution of fertilities has an infinite variance, leading to
an anomalous scaling of offsprings and generation numbers.
While the number of direct aftershocks per mainshock 
for a fixed mainshock magnitude has a finite
variance, usually modeled by a Poisson distribution, the effect of multiple
cascades of triggering 
and the variability of the fertility of each earthquake
lead to a much larger power-law distribution
for the total number of aftershocks summed over all generations.
As a consequence, there are huge fluctuations of the total number
of aftershocks from one sequence to another one, for the same 
mainshock magnitude.

The goal of this paper is to provide a general exact derivation
of the distribution of the total aftershock productivity of any 
earthquake, summed over all generations of the cascade of
aftershock triggering. We also derive the exact distribution of the
total number of generations of aftershocks before
extinction in the case of a power-law distribution of
fertility relevant for earthquakes and for many other systems.
Beyond earthquakes our results apply to any branching process with
a power-law distribution of fertilities.

\section{Model and main results}

We consider a general branching process in which each progenitor or mother is
characterized by its average number $N_m \equiv \kappa \mu(m)$ of children 
(first generation offsprings), 
where $\mu(m) = 10^{\alpha (m-m_0)}$ is a mark associated with an
earthquake of magnitude $m \geq m_0$,  $\kappa$ is a constant factor
and $m_0$ is the minimum magnitude of earthquakes capable of triggering
other earthquakes.
We note $\mu$ the mark of an earthquake which has an arbitrary magnitude $m$
drawn according to the GR law.
According to (\ref{pojid}), the mark $\mu$ is distributed according to 
\be
p_{\mu}(\mu) = {\gamma \over \mu^{1+\gamma}}~, 
~~~1 \leq \mu < +\infty, ~~~~~\gamma = b/\alpha~.
\label{aera}
\ee
Note that $p_{\mu}(\mu)$ is normalized: $\int_1^{+\infty} d\mu ~p_{\mu}(\mu)=1$.
The relation $N_m \equiv \kappa \mu(m)$ together with (\ref{aera}) thus ensures
that the fertility obeys the  law (\ref{aerapri}). 
For a fixed $\gamma$, the coefficient $\kappa$ then controls the value of the 
average number $n$ of children of first generation per mother:
\be
n = \langle N_m \rangle = \kappa \langle \mu \rangle = \kappa {\gamma \over \gamma -1}~,
\label{mgmlele}
\ee
where the average $\langle N_m \rangle$ is taken over all mothers' magnitudes
drawn in the GR law.
In the terminology of branching processes, $n$ is called the branching ratio.
For $n<1$, there are less than one child per mother: 
this corresponds to transient (sub-critical) branching processes
with finite lifetimes with probability one. For $n>1$,
there are more than one child per mother: this corresponds
to explosive (super-critical) branching processes with a number of events
growing exponentially with time. The value $n=1$ of exactly one child per mother
on average is the critical point separating the two regimes.

The realized number of children 
of an earthquake of fixed magnitude $m$
can be deterministic or may result from a Poisson or other more general
distribution with mean $\kappa \mu(m)$. Any given child $i$
may then generate an average number $\kappa \mu_i$ of children, where
the mark $\mu_{i}$ is specific to the child $i$ and is drawn from the PDF
(\ref{aera}). These grand-children of the initial progenitor in turn generate new
children with marks drawn from (\ref{aera}), and so on. The process
cascades down along generations.

Here, we focus on global quantities $p_r(r)$ (total number $r$ of offsprings
until extinction following a mainshock)
and $p_g(g)$ (total number $g$ of generations until extinction following a mainshock).
Therefore,
arbitrary time-dependent branching processes can be considered and our
results thus apply to arbitrary stochastic marked point-processes in
discrete or continuous time. In particular, our results apply 
to the ETAS model. We have previously
shown that, for $n<1$, the infinite variance of the number of first-generation
events leads to anomalous global direct and inverse Omori law with 
apparent exponents varying continuously with $\gamma$ \cite{foreshockp}.
Our results also apply to variations of the ETAS models with different 
arbitrary time dependences.

In this work, we present an analytical derivation of the distribution
of the total number of aftershocks, summed over all generations
of the cascade of aftershock triggering, and of the distribution of
the  total number of generations of aftershocks before extinction.
Specifically, we uncover a novel regime with continuously varying
exponents for the probability density
function (PDF) $p_r(r)$ of the total number $r$ of progenies and
$p_g(g)$ of their total number $g$ of generations of aftershocks 
of a mainshock before extinction. This regime appears in the regime 
of large deviations $\gamma <2$ relevant for earthquakes, 
when the distribution of the number of first-generation 
offsprings from any mother has a power law tail with infinite variance. 

\begin{figure}[h]
\includegraphics[width=8cm]{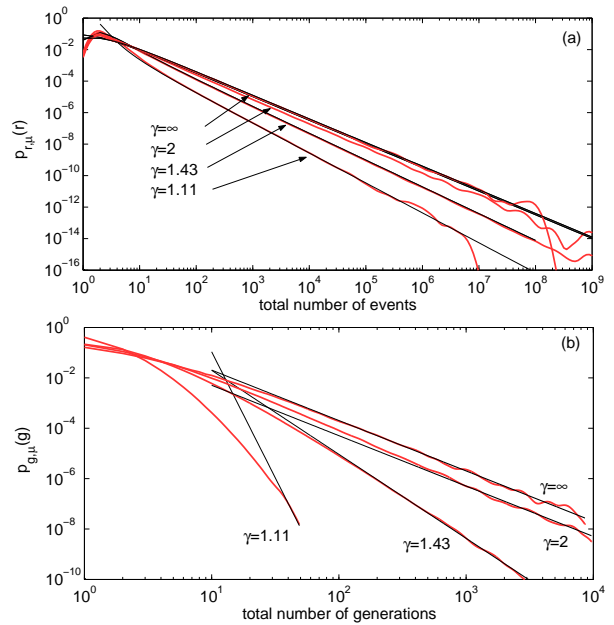}
\caption{\label{Fig1new} Numerical tests of
the anomalous scalings (\ref{affhg}) at
criticality $n= \kappa \langle \mu \rangle =1$.
The noisy lines are the Monte-Carlo simulations. The smoothed lines
are the predictions (\ref{affhg}). The deviation between
theory and numerical simulations on $P_g(g)$ for $\gamma =1.1$
results from the very large theoretical exponent equal to $1+{1 \over \gamma -1} =11$,
describing a fast fall-off which is hard to document numerically since one decade
in the horizontal scale corresponds to 11 decades on the vertical scale.}
\end{figure}

We study the sub-critical and critical
regimes for which the number $n$ of children per mother, averaged
over all possible numbers of children per mother,
is less than or equal to $1$. This condition $n \leq 1$ ensures
that, with probability $1$, the cascade ends after a finite number of
generations with a finite total number of offsprings \cite{Athreya,Sankaranarayanan}.
Figure \ref{Fig1new} presents comparisons between numerical simulations
of branching processes for different values of $\gamma$ and our main
result (at criticality $n=1$) derived below:
\be
p_r(r) \approx C_r/r^{1+{1 \over \gamma}}~~;~~
p_g(g) \approx C_g/g^{1+{1 \over \gamma -1}}~,
\label{affhg}
\ee
for $1 < \gamma \leq 2$.  
$C_r$ and $C_g$ are two constants independent of $r$ and $g$, respectively,
such that $p_r(r)$ and $p_g(g)$ are normalized to 1.
For $n<1$, the intermediate asymptotics (\ref{affhg}) holds for 
$r< R_2 \simeq 1/(1-n)^{\gamma/(\gamma -1)}$
beyond which there is another power-law asymptotic
\be
p_r(r) \sim 1/r^{1+\gamma}~.
\label{affhg2}
\ee

For $\gamma \geq 2$ and $n=1$, the standard mean field scaling
\be
p_r(r) \sim 1/r^{3/2}~~~~~~{\rm and}~~~~~~ p_g(g) \sim 1/g^2
\label{fmla}
\ee
are retrieved.
The regime $0 < \gamma \leq 1$
gives rise to explosions and, in continuous time, to stochastic
finite-time singularities \cite{sorhelmsing}.
If the GR law is truncated or exhibits a deviation from its standard
form in the large magnitude range, our results will hold
for intermediate values of $r$ and $g$ and will be
truncated at large $r$'s and large $g$'s. 

The following section gives the technical derivation of the results
(\ref{affhg},\ref{affhg2},\ref{fmla}).

\section{General formulation}

\subsection{Formal solution in terms of Probability Generating Functions (PGF)}

The most general relations of branching theory are expressed via probability 
generating functions (PGF) defined by 
\be
\langle z^R\rangle=\sum_{r=0}^\infty p_r(r)~z^r~
\ee
where $\langle\dots\rangle$ means statistical averaging. 
By definition, $p_r(r)=\text{Pr}(R=r)$ is the probability that
the random variable $R$ takes the value $r$.
It can be obtained from its PGF by the relation
\be
p_r(r)=\text{Pr}(R=r)=\frac{1}{r!}\left. \frac{d^r
\langle z^R\rangle}{dz^r}\right|_{z=0}\,.
\ee
The key property of PGFs is that the PGF of the sum of
statistically independent summands is equal to the product of the
summands PGF's. This is useful for branching processes for
which the number of daughters triggered by different mothers
are statistically independent.
We introduce four PGF's: (i) $G_{\mu_0}(z)$ is the
PGF of the number $R_1$ of daughters generated from a given
mother with fertility $\kappa \mu$ at the first generation; (ii) $G(z)$ is the
same as $G_{\mu}(z)$ but averaged over all mothers' fertility $\kappa \mu$; (iii)
$\Theta_{\mu}(z)$ is the PGF of the total number
of daughters generated from a given
mother with fertility $\kappa \mu$ summed over all generations;
(iv) $\Theta(z)$ is the same as $\Theta_\mu(z)$
but for a mother of arbitrary fertility. Then, general
branching theory \cite{Athreya,Sankaranarayanan} gives
the functional equations
\be
\Theta(z)=G(z\Theta(z))~~~;~~~\Theta_\mu(z)=G_\mu(z\Theta(z)) ~.
  \label{mdcmx}
\ee
Using $\langle R \rangle = d \Theta(z)/dz|_{z=1}$ and
similarly for the other PGF, the formulas
(\ref{mdcmx}) lead to $\langle R \rangle = n/(1-n)$ for the 
number of daughters summed over all generations and
averaged over all possible mother's fertilities,
where $n$ defined by (\ref{mgmlele}) is the average number of
first-generation daughters from mothers of arbitrary fertility.  

Using Lagrange series, we transform the implicit equation
(\ref{mdcmx}) in $\Theta_\mu$ into an explicit equation.
Recall that a Lagrange series is the Taylor expansion of
the function $G_\mu(y(z,a))$ with respect to $z$, where
$y(z,a)$ is solution of the implicit equation
$y=a+z\,G(y)$. For infinitely differentiable functions
$G_\mu(y)$ and $G(y)$ in the neighborhood of $z=a$, we
have the following Taylor series 
\be
G_\mu(y)=G_\mu(a)+\sum_{k=1}^\infty \frac{z^k}{k!}\,
\frac{d^{k-1}}{d a^{k-1}}\left[G^k(a)\frac{d
G_\mu(a)}{da}\right]\,. 
\ee 
In (\ref{mdcmx}), we have $y(z,a=0)=z\,\Theta(z)$, and using the identity
$p_{r, \mu}(r) = (1/r!) d^r \Theta_\mu(z)/dz^r|_{z=0}$
allowing to recover the PDF from its generating function,
we obtain the explicit formula for the PDF $p_{r, \mu}(r)$
of the total number of daughters born from a mother with
fertility $\mu$: 
\ba
p_{r, \mu}(0) &=& G_\mu(0)   \\
p_{r, \mu}(r) &=&
\left.\,\frac{1}{r!}\,
\frac{d^{r-1}}{d z^{r-1}} \left[G^r(z) \frac{d
G_\mu(z)}{dz}\right]\right|_{z=0}~,~~~~~r>0~. \label{mvme} 
\ea

Let us now specialize to the Poisson distribution
\be
p_{r_1,\mu}(r_1)= \frac{(\mu \kappa)^{r_1}}{r_1!}\, e^{-\mu \kappa} ~,
\label{mnjjkez}
\ee
giving the PDF of the number $R_1$ of daughters of first generation
born from a mother with fertility $\mu$. By construction, the average of $R_1$
at fixed mother fertility $\mu$ over an ensemble of Poisson realizations is
$\langle R_1 \rangle_{\mu} = \mu \kappa$. The PGF associated with
(\ref{mnjjkez}) is
\be
G_\mu(z)=e^{\mu \kappa (z-1)}\,.
\label{mles}
\ee
 From (\ref{mles}), we obtain the expression of $G(z)$ by averaging
over all possible fertilities $\mu$ according to the PDF
(\ref{aera}). Using explicitly the normalized PDF
$p_{\mu}(\mu) = \gamma/\mu^{1+\gamma}$ for $\mu \geq 1$
and $0$ otherwise, we obtain 
\be
G(z)=\gamma\,\kappa^\gamma (1-z)^\gamma\,
\Gamma(-\gamma,\kappa(1-z))\,. \label{cnnbwsz} 
\ee
Expression (\ref{cnnbwsz}) can be expanded as \be
G(z)\simeq 1- a_1 ~y ~-~ a_{\gamma}~y^\gamma ~+~ a_2~y^2
~+~\dots \label{zmmlw} \ee where $y=\kappa (1-z)$,
$a_1=\frac{\gamma}{\gamma-1}$, $a_{\gamma}=
\Gamma(1-\gamma)$ and $a_2=\frac{\gamma}{2(\gamma-2)}$.
This expansion (\ref{zmmlw}), valid for any $\gamma$,
applies beyond the specific Poisson process
(\ref{mnjjkez}). It is solely based on the asymptotic
power law PDF (\ref{aera}) of the average number of
daughters at the first generation. Thus, our asymptotic
results (\ref{affhg}) derived below hold under very
general conditions.

\subsection{Probabilistic interpretation \label{probainter}}

Let us provide a probabilistic
interpretation of formula (\ref{mvme}). In addition to 
be of intuitive appeal, it will be very useful for 
the derivation of our results. 

Let us consider
the random integer $R$ with PDF $P(k)=\text{Pr}(R=k)$ and its PGF as
\be
S(z)=\sum_{k=0}^\infty P(k) z^k\,,
\ee
and reciprocally
\be
P(k)=\text{Pr}(R=k)=\frac{1}{k!}\left. \frac{d^k
S(z)}{dz^k}\right|_{z=0}\,. 
\label{dfi}
\ee
Let us decompose $R=X+Y$, where $X$ and $Y$ are
statistically independent random integers with PGFs 
$Q(z)$ and $G(z)$ respectively. 
Due to the statistical independence of $X$ and $Y$, we
have $S(z)=Q(z)G(z)$ and relation (\ref{dfi}) takes the form
\be
P(k)=\text{Pr}(X+Y=k)=\frac{1}{k!}
\left. \frac{d^k} {dz^k} \left[Q(z)G(z)\right] \right|_{z=0}\,.
\label{d}
\ee
Let in turn
\be
Y=Y(r)=Y_1+Y_2+\dots+Y_r 
\label{e}
\ee
where $\{Y_1,\dots,Y_r\}$ are $r$
statistically independent random integers with the same
PGF $G(z)$. Then, formula (\ref{d}) takes the form
\be
P(k)=\text{Pr}(X+Y(r)=k)=\frac{1}{k!}\left. \frac{d^k
}{dz^k} \left[Q(z)G^r(z)\right] \right|_{z=0}\,.
\label{f}
\ee
Let us now rewrite relation (\ref{mvme})
in a form similar to (\ref{f}). For this,
let us introduce the auxiliary PGF
\be
Q_\mu(z)=\frac{1}{\left<r_1(\mu)\right>} \frac{d G_\mu(z)}{d z}\,.
\ee
It is easy to prove rigorously that, for an arbitrary PGF
$G_\mu(z)$ possessing a finite mean value
\be
\left<r_1(\mu)\right>=\left.\frac{d G_\mu(z)}{d z}
\right|_{z=1}<\infty~,
\ee
then the auxiliary function $Q_\mu(z)$ is indeed 
the PGF of some random variable. Moreover, in the 
case under consideration of a Poissonian PGF, we have
\be
G_\mu(z)=e^{\mu\kappa(z-1)}\qquad\Rightarrow\qquad
Q_\mu(z)\equiv G_\mu(z)=e^{\mu\kappa(z-1)} 
\label{g}
\ee
and
\be
\left<r_1(\mu)\right>=\mu\kappa\,. 
\label{h}
\ee
One can thus rewrite relation (\ref{mvme}) in the form
\be
p_{r,\mu}(r)=\left.\,\frac{\mu\kappa}{r!}\, \frac{d^{r-1}}{d
z^{r-1}} \left[G^r(z) G_\mu(z)\right]\right|_{z=0}\,.
\label{i}
\ee
Taking into account relation (\ref{f}) and introducing the new
random integer $X(\kappa\mu)$ possessing the Poissonian
PGF (\ref{g}), we can rewrite (\ref{i}) in the form
\be
p_{r,\mu}(r)=\frac{\mu\kappa}{r}\,
\text{Pr}\left[X(\kappa\mu)+Y(r)=r-1\right]\,. 
\label{j}
\ee
This is the key formula that we will use for deriving the
asymptotic relations for the PDF $p_{r,\mu}(r)$ using limit theorems.

\section{Distribution of the total number of aftershocks}

\subsection{Case $\gamma \to \infty$ }

Let us first consider the limiting case $\gamma \to
\infty$. The formal limit of a power law 
with an exponent going to infinity can be assimilated to an
exponential in the following sense. Writing the tail
of the complementary distribution as $A^{\gamma}/ (A+\mu)^{\gamma}$,
and assuming that $A$ also grows with $\gamma$ as $A=\gamma/a$,
then $A^{\gamma}/ (A+\mu)^{\gamma} = 1/(1+a\mu/\gamma)^{\gamma}
\to_{\gamma \to +\infty} \exp[-a \mu]$. This 
corresponds to an exponential tail for the distribution of
the number of first generation daughters for arbitrary
mother's fertilities, as in (\ref{mnjjkez}) but with the same 
fertility for all mothers.
In this case, $n=\kappa$ as can be seen from (\ref{mgmlele}). 
This amounts to replacing
expression (\ref{cnnbwsz}) by $G(z)=e^{\kappa(z-1)}$ and we also have
$G_\mu(z)=e^{\mu\kappa(z-1)}$. Thus, 
$$
G^r(z) \frac{d G_\mu(z)}{dz}=\mu\kappa\, e^{\kappa(z-1)(r+\mu)}~.
$$
Correspondingly,
\be
\frac{d^{r-1}}{d z^{r-1}} \left[G^r(z) \frac{d
G_\mu(z)}{dz}\right]=\mu\kappa^r (r+\mu)^{r-1}\,
e^{\kappa(z-1)(r+\mu)}~.  
\label{mgmjlkled}
\ee
By substituting (\ref{mgmjlkled}) into the right-hand-side of 
(\ref{mvme}) gives exactly 
\be
p_{\mu}(r)=\mu\,\kappa^r~ e^{-\kappa \mu}\,
\frac{\left[(\mu+r)\,e^{-\kappa}\right]^r} {r!(\mu+r)}\,.
\label{mgbjjb} 
\ee 

There are two ways of deriving asymptotic formulas
for $p_{r,\mu}(r)$ in the case $\gamma=\infty$. The first one relies
on Stirling's formula $r!\simeq \sqrt{2\pi r}\, r^r e^{-r}$
applied to expression (\ref{mgbjjb}).
The second method relies on the probabilistic
interpretation of the general formula (\ref{mvme}) given in section
\ref{probainter} together with the central limit theorem.
The formulas obtained by these two methods differ
in details, but are actually very close quantitatively
over the most part of the tail, and are easily checked to converge to the same
result asymptotically. The approach using the Stirling formula
works only for $\gamma=\infty$ and does not allow to derive more
general results for $1<\gamma<\infty$. The second approach
is much more powerful and elegant. In
particular, all asymptotics and crossovers in the case
$1<\gamma<\infty$ are obtained by the second approach.

Let us first examine the first method.
Using the above mentioned Stirling's formula, one can rewrite
(\ref{mgbjjb}) in the form
\be
p_{r,\mu}(r)=\frac{1}{\mu}\,f(x)\,\qquad x=\frac{r}{\mu}\,,
\label{a}
\ee
where
\be
f(x) \simeq \frac{C}{\sqrt{2\pi x}~(1+x)}
\left(1+\frac{1}{x}\right)^{\mu x}\,e^{-\mu\delta x}
\label{b}
\ee
and
\be
C=\sqrt{\mu}\,e^{-\kappa\mu}\,,\qquad
\delta=\kappa-\ln\kappa-1\,.
\label{c}
\ee
Because Stirling's formula is very precise even for
$r\simeq 1$ (for example $1!=1$ while Stirling's formula
gives $1!\simeq0.922$, $2!=2$ while Stirling's gives
$2! \simeq 1.919$ and so on),
formula (\ref{b}) actually works well even for not too large
$\mu$ and intermediate $\kappa<1$. 
The advantage of (\ref{b}) is that it gives a clear
understanding of the structure of the PDF $f(x)$. It consists of
(i) a characteristic power tail
\be
f(x)\propto \frac{1}{\sqrt{2\pi x}(1+x)}\sim x^{-3/2}\,,
\ee
(ii) an exponential decaying factor
\be
f(x)\propto e^{-\mu\delta x}
\ee
which disappears when $\kappa=1$, i.e., $\delta=0$, and
(iii) an algebraic factor
\be
f(x)\propto \left(1+\frac{1}{x}\right)^{\mu x}\sim
e^\mu\, \exp\left(-\frac{\mu}{x}\right)
\ee
which possesses a lower cut-off at $x\lesssim \mu$.

The shortcoming of (\ref{b}) is that we can derive it
only when an explicit expression such as (\ref{mgbjjb}) is
available. We have not such luxury in the more general case
$1<\gamma<\infty$ for which we need the second
approach. Let us first quote the asymptotic formula (\ref{hgfrw})
given below for $\gamma \to +\infty$ obtained from the 
application of the second probabilistic method. We shall 
show how to derive (\ref{hgfrw}) as a special case
of the next section for $2 < \gamma$.

When the variance $\sigma_1^2$ of the
number $r_1$ of daughters of the first generation from
mothers of arbitrary fertilities is much greater than
$1-\kappa$, we find (see next section) that, due to central limit theorem,
expression (\ref{mgbjjb}) reduces to
\be 
p_{\mu}(r)
\simeq \frac{\mu\kappa}{r\sqrt{2\pi\,\kappa\,(r+\mu)}}\,
\exp\left(-\frac{[(1-\kappa)r-\mu\kappa-1]^2}
{2\kappa\,(r+\mu)}\right)\,.  
\label{hgfrw} 
\ee

At criticality, $n=\kappa=1$, this expression becomes
 \be
p_{\mu}(r) \simeq \frac{\mu}{r\sqrt{2\pi r}}\,
\exp\left(-\frac{\mu^2}{2 r}\right)\,, 
\label{mgjsz} 
\ee
which retrieves the announced well-known mean field
asymptotics $p_{\mu}(r) \sim 1/r^{3/2}$.
The exponential term $\exp(-\mu^2/(2 r))$ describes the roll-off
of the number of aftershocks for small numbers close to the
characteristic value $r \approx \mu^2$.

In the subcritical regime $n<1$, expression (\ref{mgbjjb} ) gives an
exponential decay of the number of aftershocks for large $r$. 

These results can be checked explicitly as follows.
The case of $\gamma \to +\infty$ corresponds either to $b \to \infty$
(all events have the same magnitude) or to $\alpha=0$ (all the events trigger
the same expected number of offsprings independently of their magnitude). 
In both cases, $\mu$ tends to a constant independent of the mainshock 
magnitude. 
In order to check the above results, let us take this constant equal to $1$. This 
choice will modify the constants but not the functional dependences.
Assuming $\mu=1$ transforms (\ref{mgbjjb}) into
Taking  $\mu=1$ transforms (\ref{mgbjjb}) into
\be
p_{\mu}(r)= \kappa^r ~{(r+1)^r \over (r+1)!}~e^{-\kappa (r+1)}~.
\label{mgbaaajjb} 
\ee 
By using the Stirling formula, it has the approximation
\be
p_{\mu}(r) \approx { e^{r \ln \kappa + (r+1)(1-\kappa)} \over \sqrt{2\pi}~(r+1)^{3/2}}~.
\ee
When $\kappa=n=1$ (critical case), we retrieve $p_{\mu}(r) \sim r^{-3/2}$.
When $\kappa = n < 1$ is close to $1$,
$p_{\mu}(r) \sim r^{-3/2}~e^{-\delta ~r}$
where $\delta$ is defined in (\ref{c}).

\subsection{Case of finite variance $2 < \gamma < \infty$}

In this case, the summands in the sum (\ref{e}) have
a finite mean $\left<Y_1\right>$ and finite variance
$\sigma^2_1$. Thus, the Central Limit Theorem (CLT) holds
(see for instance \cite{Sorbook} for a pedagogical exposition
of the CLT). Let us recall the explicit expressions of the mean
and variance:
\be
\left<Y_1\right>=n=\frac{\kappa\gamma}{\gamma-1}\,,
\qquad \sigma_1^2=\frac{n^2}{\gamma(\gamma-2)}+n\,,
\label{k}
\ee
If $r\gg 1$ (in practice, it is sufficient that $r\gtrsim 6$), 
the sum $Y(r)$ in (\ref{e})
converges to a Gaussian variable with the following
mean and variance
\be
\left<Y(r)\right>=r\,n\,,\qquad \sigma^2(r)=r \sigma^2_1\,. 
\label{l}
\ee

Let $\mu\kappa$ be integer (just for illustrative
purpose). Then, one can imagine $X(\mu\kappa)$ as a sum
\be
X(\mu\kappa)=\sum_{m=1}^{\mu\kappa} X_m 
\label{m}
\ee
of independent summands
$\{X_1,X_2,\dots,X_{\mu\kappa}\}$ with the following
PGF, mean and variance:
\be
Q(z)=e^{z-1}\,,\qquad \left<X_1\right>=\sigma^2=1\,.
\ee
If the number of summands $\mu\kappa\gg 1$, then the sum (\ref{m}) is
asymptotically Gaussian as well, with mean and variance
equal to $\mu\kappa$.

Thus if $r\gg 1$ and $\mu\kappa\gg 1$, then the sum
$X(\mu\kappa)+Y(r)$ is asymptotically Gaussian and we
can use the following asymptotic formula
\be
\text{Pr}\left[X(\kappa\mu)+Y(r)=m\right]=
\frac{1}{\sqrt{2\pi\sigma_\mu^2(r)}} \exp\left[ -
\frac{(m-a_\mu(r))^2}{2\sigma_\mu^2(r)} \right] \,.
\label{n}
\ee
Here $a_\mu(r)$ and $\sigma_\mu^2(r)$ are respectively the
mean and variance of the sum $X(\mu\kappa)+Y(r)$:
\be
a_\mu(r)=r n+\mu\kappa\,, \qquad \sigma_\mu^2(r)=r
\sigma_1^2+\mu\kappa\,. 
\label{o}
\ee
Substituting (\ref{o}) into (\ref{n})
and (\ref{n}) into (\ref{j}), we obtain 
\be
p_{r,\mu}(r)\simeq \frac{\mu\kappa}{r\sqrt{2\pi(\sigma_1^2
r+\mu\kappa)}}\,
\exp\left(-\frac{[(1-n)r-\mu\kappa-1]^2} {2(\sigma_1^2
r+\mu\kappa)}\right)\,. 
\label{p}
\ee

For $\gamma=\infty$, for which we have the mean and variance
defined in (\ref{k}) reduce to $\sigma^2_1=n=\kappa$,
relation (\ref{p}) transforms into the previously
announced result (\ref{hgfrw}). In this case, we can test the
quality of the general asymptotic relation (\ref{p})
by comparing its particular application (\ref{hgfrw}) with 
the exact expression (\ref{mgbjjb}) 
and its Stirling asymptotics (\ref{b}). 
The exact and Stirling's approximation essentially coincide while
the ``Gaussian''
approximation is also excellent and goes closer and closer
to the exact formula the larger $\mu$ is and the closer
$n=\kappa$ is to $1$.

For $\sigma_1^2~ r > \mu\kappa$, expression (\ref{p}) can be simplified into
\be
p_{\mu}(r) \simeq
\frac{\mu\kappa}{r\sqrt{2\pi\sigma_1^2\, r}}\,
\exp\left(-\frac{[(1-n)r-\mu\kappa-1]^2} {2\sigma_1^2\,
r}\right)\,,  
\label{mgmlelwe}
\ee
which reduces again to the standard mean field asymptotics
(\ref{mgjsz})
at the critical point $n=1$ and for $r \gg \mu^2\kappa^2$.
Expression (\ref{mgmlelwe}) is obtained using the 
following approximation
\be
\sigma_1^2 r+\mu\kappa\simeq \sigma_1^2 r\,, 
\label{q}
\ee
that is, by neglecting the term $\mu\kappa$. 
From a probabilistic point of view, this corresponds
to using the Law of Large Numbers for the sum (\ref{m}). 
Namely, if $\mu\kappa \gg 1$, 
then one can replace (\ref{m}) by its mean field limit
\be
X(\mu\kappa)\simeq \left<X(\mu\kappa)\right>=\mu\kappa~,
\label{r}
\ee
which provides the truncated equality (\ref{q}). Another reason for
using the mean field limit (\ref{r}) is more technical:
in order to be able to  effectively use the
same technique in the case
$1<\gamma<2$, we need to implement such a ``mean field approximation.''
Otherwise, the formulas would include
very complicated convolutions of Gaussian
and Levy stable laws.

We have checked the validity of the mean field approximation by
comparing numerically the general asymptotics (\ref{p}) and of its
truncated mean field limit version (\ref{mgmlelwe}) for various
values of $\gamma$,  $n$ and $\mu\kappa$. We find always
an excellent convergence in the tail. For instance, consider the 
simplest case $\gamma=\infty$, for which the general
asymptotic formula (\ref{p}) transforms into (\ref{hgfrw}), while
its truncated mean field version is
\be
p_{r,\mu}(r)\simeq \frac{\mu\kappa}{r\sqrt{2\pi\kappa r}}\,
\exp\left(-\frac{[(1-\kappa)r-\mu\kappa-1]^2} {2\kappa r}\right)\,. 
\label{s}
\ee
The agreement between all these expressions is excellent in the tails
for all values of the parameters. The truncated mean field approximation
describes satisfactory the body of distribution and becomes very precise
in the tail of the distribution, even for moderate values
of $\mu \approx 25$ and $n=\kappa <0.9$. Even better agreement is obtained
for larger $\mu$ and closer to the critical case $n=1$.

Let us complement this section by a few 
additional useful formulas. The expansion (\ref{zmmlw}) with
$\langle r_1 \rangle = dG(z)/dz|_{z=1}$ shows that the
average number $r_1$ of first-generation daughters from mothers
of arbitrary fertilities is $n=\langle \mu \rangle \kappa = \kappa \gamma
/(\gamma -1)$
and that, for $1 < \gamma < 2$,
the variance of $r_1$ is infinite. For $\gamma > 2$, the variance of
$r_1$ is $\sigma_1^2=n + n^2/[\gamma(\gamma -2)]$. The form of (\ref{zmmlw})
is essentially controlled by the expression of the
characteristic function of the PDF (\ref{aera}) and it is thus
not a surprise that the PDF of the number of daughters of
first generation born from a mother with arbitrary fertility
has the same asymptotic form as (\ref{aera}).
For a mother with fixed fertility $\mu$, the average and variance
of the total number of daughters are respectively
\be
\langle r_{\mu} \rangle= \frac{\mu \kappa}{1-n}~~;~~
\sigma^2_{\mu}=\frac{\mu \kappa}{(1-n)^3}
\left[1+\frac{n^2}{\gamma(\gamma-2)}\right]~,~~~~{\rm for}~~\gamma >2~.
\ee

\subsection{Case of infinite variance $1 < \gamma < 2$}

We now turn to the novel regime $1<\gamma<2$. In this case,
it is convenient to rewrite relation (\ref{mvme}) in a more
transparent probabilistic form 
\be
p_{\mu}(r)=\frac{\mu\kappa}{r}\,
\text{Pr}\left[X+Y(r)=r-1\right]\,, 
Y(r)=\sum_{k=1}^r Y_k\,, \label{probabil} 
\ee 
where $\{X, Y_1,\dots, Y_r\}$ are mutually independent random integers
such that the PGF of $X$ is $G_\mu(z)$ given by (\ref{mles}) while the PGF's
of the remaining integers are equal to $G(z)$ given by (\ref{cnnbwsz}). For
$1<\gamma<2$, the variance of each variable $Y_k$ is
infinite, for the same reason that $\sigma_1^2$ is
infinite. From the generalized limit theorem \cite{Major},
$\tilde{Y}(r)=Y(r)-nr$ converges in distribution to a PDF
$f_y(s)$ which is proportional to a stable
infinitely-divisible PDF $\varphi_\gamma(x)$ with exponent
$\gamma$ 
\be
f_y(s)={ 1 \over \nu r^{1/\gamma}} \varphi_\gamma\left(
{ s \over \nu r^{1/\gamma} } \right)~,
\label{f2}
\ee
where 
\be
\nu=\kappa[\gamma\Gamma(-\gamma)]^{1/\gamma},
\label{nu}
\ee
and
\be
\varphi_\gamma(x)=\int \limits_0^{\infty} 
\exp \left(u^{\gamma} ~\cos \left({ \pi~\gamma \over 2}\right)\right)~
    \cos \left(u^\gamma~\sin({ \pi~\gamma \over 2})+u~x \right)~du
\label{levy}
\ee
with  $\varphi_\gamma(0)= -1/\Gamma\left(-1/\gamma\right)$.
Thus we obtain the distribution of aftershock number from 
(\ref{probabil}) and (\ref{f2})
\be 
p_{\mu}(r) \simeq
\frac{\mu\kappa}{\nu r^{1+{1 \over \gamma}}}\,
\varphi_\gamma\left(\frac{(1-n)r-\mu\kappa-1}{\nu
r^{1/\gamma}} \right)\,. 
\label{mgmmss} 
\ee

For large number of aftershocks $ r>>R_1$ where $R_1$ is defined by
\be
R_1 = \left(\frac{\mu\kappa+1}{\nu}\right)^\gamma\simeq
\frac{\mu^\gamma}{\gamma\Gamma(-\gamma)}~, 
\label{R1}
\ee
we can use the following asymptotic of $\varphi_\gamma(x)$ for $x \to \infty$
\be
\varphi_\gamma(x)\sim \frac{x^{-1-\gamma}}{\Gamma(-\gamma)}
\label{opih}
\ee
Putting (\ref{opih}) in (\ref{mgmmss}) retrieves at criticality ($n=1$)
our main result announced in (\ref{affhg}) with
$C_r = \mu\,\varphi_\gamma(0)/[\gamma\Gamma(-\gamma)]^{1/\gamma}$.

\begin{figure}[h]
\includegraphics[width=8cm]{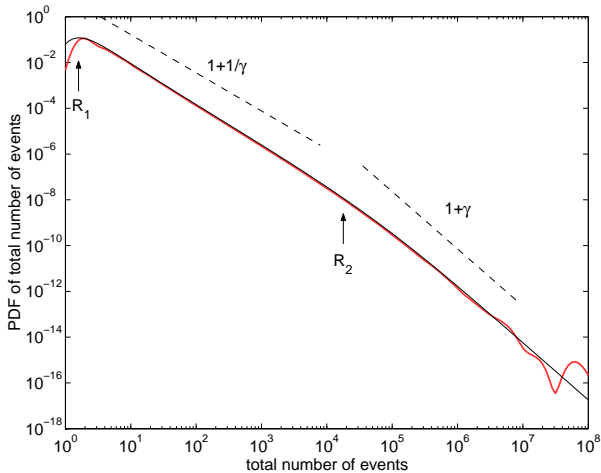}
\caption{\label{Fig2} Numerical tests of the cross-over
between (\ref{affhg}) and (\ref{eq31}) which is
predicted by (\ref{mgmmss}) of the PDF $p_{\mu}(r)$.
The thick (resp. thin) line
is the numerical simulation (resp. prediction (\ref{mgmmss}))
for $n=0.97$ and $\gamma=1.5$. The two
dashed lines outline the first intermediate asymptotic with
exponent $1+1/\gamma$ for $R_1 \ll r \ll R_2$ and truly asymptotic 
with exponent $1+\gamma$ for $r \gg R_2$. 
The crossovers $R_1=1.5$ and $R_2=1.6\times 10^4$ are defined by (\ref{R1}) 
and (\ref{R2}) respectively.
}
\end{figure}

In the subcritical case ($n<1$), there is another power
asymptotics. For $r\gg R_1$, one can rewrite (\ref{mgmmss}) in the form
\be
p_{\mu}(r) \simeq
\frac{\mu}{[\gamma\Gamma(-\gamma)]^{1/\gamma} r^\beta}\,
\varphi_\gamma\left(\frac{1-n}{\nu}
r^{\frac{\gamma-1}{\gamma}} \right)\,.
\label{eq29}
\ee
If additionally $r\gg R_2$, where $R_2$ is defined by  
\be
R_2 =\left(\frac{\nu}{1-n}\right)^{\gamma/(\gamma-1)}~,
\label{R2}
\ee
then, using the asymptotic (\ref{opih}) of $\varphi_\gamma(x)$, we obtain
\be
p_{\mu}(r) \simeq \mu\,\gamma\,
\left(\frac{n(\gamma-1)}{(1-n)\gamma}\right)^{\gamma+1}\,
~~{1 \over r^{1+\gamma}}\,,
\label{eq31}
\ee
for $r\gg \text{max}\{R_1,R_2\}$.
If additionally $R_2 \gg R_1$, which holds if
$\frac{1-n}{n}\ll (\gamma-1)\Gamma(-\gamma)\,\mu^{1-\gamma}$,
then for $R_1\ll r\ll R_2$, there is an intermediate
asymptotics following expression (\ref{affhg}). These results
are checked in Figure \ref{Fig2}.

\section{Distribution of the total number of generations}
We now turn to the determination of the PDF of the total number
of branching generations, in other words, to the probability
that the branching process terminates at a given generation number.
Let $U_\mu(z;g)$ (respectively $U(z;g)$) denote the PGF corresponding 
to the number of daughters born from a mother with fertility $\mu$
(respectively with arbitrary fertility)
at the $g$-th generation. A standard result of branching theory is \cite{Athreya}
\be 
U_\mu(z;g+1)=G_\mu(U(z;g))~;~U(z;g+1)=G(U(z;g)),
\label{eq1bis}
\ee
where $U(z;0)=z$. The probability that the branching
process survives at the $g$-th generation is
\be
p_\mu(g)=1-U_\mu(0,g)~.
\label{eq3bis}
\ee
The probability of termination of the branching
process at the $g$-th generation is then given by
\be
p_{g,\mu}(g)=p_\mu(g)-p_\mu(g+1)\,.
\label{eq4bis}
\ee
 From (\ref{eq1bis},\ref{eq3bis}), the probability for
the branching process to survive at the
$g$-th generation is
\be
p_\mu(g+1)=1-G_\mu(1-p(g))\,.
\label{eq5bis}
\ee
$p(g)$ is the surviving probability at the $g$-th generation
triggered by a mother of arbitrary fertility $\mu$, which obeys
the recurrence
\be
p(g+1)=1-G(1-p(g))\,,\qquad p(0)=1\,.
\label{eq6bis}
\ee

Equations (\ref{eq5bis}) and (\ref{eq6bis}) are
easily solved numerically. We now extract the asymptotic
power law behavior of $p_{g,\mu}(g)$.
For small $p(g)$, the right-hand-side of (\ref{eq6bis}) can
be expanded as
\be
1-G(1-p) \simeq n p-\frac{C}{\gamma-1}p^\gamma-D ~p^2\,,
\label{eq7bis}
\ee
where $C=n^\gamma\,\Gamma(2-\gamma) [(\gamma-1)/\gamma]^\gamma$ and
$D=n^2\,(\gamma-1)^2/[2\gamma(\gamma-2)]$. For $\gamma>2$, it is enough to
take the leading behavior of (\ref{eq7bis}) up to second order
in powers of $p$: $1-G(1-p)\simeq n p-D p^2$. With
(\ref{eq6bis}), this gives $p(g+1)-p(g)=-(1-n)p(g)-Dp^2$.
Close to criticality $n \to 1$ for which a large number $g$
of generations occur, the leading behavior of the survival
probability can be obtained by taking a continuous approximation
to (\ref{eq6bis}), giving
$\frac{d p(g)}{dg}=-(1-n)p-D~p^2$.
Its solution with initial condition $p(g_0)=p_0$ has the form
\be
p(g)=\frac{p_0 (1-n)}{\displaystyle(1-n+p_0 D)\,
e^{(1-n)(g-g_0)}-p_0 D}\,.
\label{eq12bis}
\ee
Expression (\ref{eq12bis}) gives a power law $p(g) \simeq 1/(D~g)$ for
$1 \ll g < 1/(1-n)$ and crosses over to an exponential law for
$g \geq 1/(1-n)$.
Knowing $p(g)$, $p_\mu(g)$ given by (\ref{eq5bis}) can be obtained.
For instance, in the case of a Poisson distribution (\ref{mnjjkez})
giving (\ref{mles}), we obtain $p_\mu(g+1)=1- \exp\left[-\mu\,\kappa p(g)\right]$.
For $\mu \kappa\,p(g)\ll 1$, $p_\mu(g+1)$ becomes
$p_\mu(g+1)\simeq \mu \kappa\,p(g)$.
The sought PDF of $p_{g,\mu}(g)$ defined by (\ref{eq4bis})
is given by
\be
p_{g,\mu}(g)\simeq \mu \kappa\, {d p(g) \over dg} \simeq  {\mu \kappa \over D ~g^2}.
\ee
The last equality holds for $g \geq 1/(1-n)$. This retrieves
the standard mean field asymptotics $p_{g,\mu}(g) \sim 1/g^2$.

For $1<\gamma<2$, it is sufficient to keep only the following terms
$1-G(1-p)\simeq n p-\frac{C}{\gamma-1}p^\gamma$ in the expansion
(\ref{eq7bis}). Taking again a continuous approximation of (\ref{eq6bis}) gives
$\frac{d p(g)}{dg}=-(1-n)p-\frac{C}{\gamma-1} p^\gamma$,
whose solution, for $g \ll  1/[(\gamma-1)(1-n)]$, reads
\be
p(g)\simeq \left[p_0^{1-\gamma}+ C(g-g_0) \right]^{-1/(\gamma-1)} \sim
{ C \over (g-g_0)} ^{1/(\gamma-1)}~,
\label{eq23bis}
\ee
Correspondingly, the PDF $p_{g,\mu}(g)$ that the number of generations is
exactly $g$ is given by (\ref{affhg}) with
$C_g = \frac{\mu n}{\gamma}\, C^{-1/(\gamma-1)}$, valid for $\gamma$ not too close to $2$.

\section{Concluding remarks}

We have shown that the existence of cascades of triggered seismicity 
produces huge fluctuations of the total number of aftershocks and of
the total number of generations from one sequence to another one
for the same mainshock magnitude,
characterized by power-law asymptotics. 
If the distribution of offsprings per mothers 
(of direct aftershocks per mainshock for seismicity)
has a finite variance, 
we recover the well known mean-field results (\ref{fmla}). 
In the case of an infinite variance relevant for earthquakes, 
we have discovered a new regime
with exponents that varies continuously with the exponent $\gamma=b/\alpha$.
The anomalous scaling reported here for $1< \gamma <2$ gives rise
to less wild fluctuations in the total
number of daughters from one mother to the next, compared with the
mean field regime $\gamma \geq 2$. For instance, for earthquakes, we have
probably $\gamma \approx 1.25$ 
for the preferred values $b \approx 1$ and $\alpha \approx 0.8$
\cite{alpha}, which leads
to $p_r(r) \sim 1/r^{1+0.8}$ and $p_g(g) \sim 1/g^{1+4}$ at $n=1$,
compared with $p_r(r) \sim 1/r^{1+0.5}$ and $p_g(g) \sim 1/g^{1+1}$
for $\gamma \geq 2$. The reason for this behavior lies in the
important role exerted by the
mothers with the largest fertilities $\propto \mu$ on the rate of daughter births. 
In particular, this role
explains why, away from criticality but close to it ($n \leq 1$), $p_r(r)$
crosses over from $1/r^{1+{1 \over \gamma}}$ to $1/r^{1+\gamma}$
as $r$ increases:
the latter asymptotic regime is nothing but the tail of the PDF (\ref{aera})
which controls the number of daughters born in the first generation
from a mother with arbitrary fertility.

How relevant is the infinite variance regime $\gamma < 2$ in view of the uncertainties
in the exponents $\alpha$ and $b$? Indeed, 
the value of $\alpha$ is not only
approximately known but its constancy (as a function
of time) and uniformity (as a function of space) remains to be ascertained. In addition, the
$b$-value also fluctuates and is typically in the range $0.75-1.25$
\cite{Frohlich}. If we take the largest $b$-value of this interval ($b=1.25$),
we remain in the infinite variance regime $\gamma < 2$ as long as $\alpha > 0.63$.
Most present estimations discussed in section 2
give values above this lower bound. It thus seems that
the results presented here should remain pertinent as more precise estimations
of $\alpha$ and $b$ become available.

All these discussions assume untruncated power law distributions. However, it is 
well-known that the Gutenberg-Richter law exhibits an upper magnitude cut-off
\cite{Kagan11,Pisor22}. By the logic leading to (\ref{aerapri}), this 
automatically implies also an upper fertility cut-off. This in turn removes
the divergence of the variance for $\gamma < 2$. How does this affect
the distributions $p_r(r)$ and $p_g(g)$? This question is standard 
in the theory of power laws (see for instance \cite{Sorbook} and references therein).
Since the upper cutoff is very large so that the actual range of events is very
large, our results obtained without truncation hold for a large range of 
values of $r$ and of $g$ respectively. However, the predicted power laws (\ref{affhg})
have to cross-over to the mean-field ones (\ref{fmla}) at threshold values
beyond which the finiteness of the variance of fertilities become flagrant.
In contrast, the presence of a lower cutoff magnitude \cite{benzion} does
not modify our results on the power law tail and has an influence only in shaping
the bulk (small values) of the distributions.

The direct validation of our predictions (\ref{affhg}) and (\ref{affhg2}) 
on earthquake catalogs is not feasible, due to the impossibility to
distinguish aftershocks from uncorrelated events at large times after
the mainshock, and due to the limited number of large aftershock sequences. 
Our results have however some consequences for the
statistical properties of aftershock sequences. We have indeed
shown in \cite{Bathlawpap} that the existence of large fluctuations of
the number of aftershocks per mainshock (summed over all generations)
induces a non-trivial scaling of the difference in magnitude between a
mainshock and its largest aftershock, so that B{\aa}th law can be
recovered in the regime $b \gtrsim  \alpha$.
These results also suggest that the common use of a Poisson distribution
in seismicity forecasts to model the distribution of the
number of events within a finite space-time window is questionable,
since the simple physics of cascades of earthquake triggering gives
a very different (power law) distribution. Recent observations show
non-Poisson power law distributions of seismic rates
(see \cite{PG96} and work in progress). We suggest that these
observations and our results could be used to improve earthquake
forecasting by
providing a more realistic distribution of the number of events.

 \vskip 0.5cm
{\bf Acknowledgments:}  We are grateful to two unknown referees and to
Y. Ben-Zion as the editor, for useful remarks that helped improve
the manuscript.
This work is partially supported by NSF-EAR02-30429, by
the Southern California Earthquake Center (SCEC) and by
the James S. Mc Donnell Foundation 21st century scientist
award/studying complex system. 
SCEC is funded by NSF Cooperative Agreement EAR-0106924 and USGS Cooperative
Agreement 02HQAG0008.  The SCEC contribution number for this paper is 741.

\end{document}